\documentstyle{article}

\font\tenrm=cmr10
\font\tenit=cmti10
\font\elevenbf=cmbx10 scaled\magstep 1
\font\elevenrm=cmr10 scaled\magstep 1
 1

\renewenvironment{thebibliography}[1]
 { \elevenrm
   \begin{list}{\arabic{enumi}.}
    {\usecounter{enumi}     \setlength{\parsep}{0pt}
     \setlength{\itemsep}{3pt} \settowidth{\labelwidth}{#1.}
     \sloppy
    }}{\end{list}}

\begin{document}
\begin{center}{{\elevenbf
          CHARACTERISTIC PREDICTIONS OF TOPOLOGICAL SOLITON MODELS}\\}
\vglue 1.0cm
{\tenrm V.B. Kopeliovich \\}
{\tenit Institute for Nuclear Research of the Russian Academy of
Sciences,\\ 60th October Anniversary Prospect 7A, Moscow 117312, Russia\\}
\end{center}
\vglue 0.3cm
{\rightskip=2pc
 \leftskip=2pc
\tenrm\baselineskip=12pt
 \noindent
The characteristic predictions of chiral soliton models - the Skyrme model and
its extentions - are discussed. 
The chiral soliton models predictions of low-lying dibaryon states qualitatively
agree with recent evidence for the existence of narrow dibaryons 
in reactions of inelastic proton scattering on deuterons and double photon
radiation $pp \rightarrow pp \gamma \gamma$.
The connection between magnetic momentum operators and tensors of inertia
valid for arbitrary $SU(2)$ skyrmion configurations allows to estimate the 
electromagnetic decay width of some states of interest. 
Another kind of predictions are multibaryons with nontrivial flavour -
strangeness, charm or bottom, which can be found, in particular, in high energy
heavy ions collisions. 
It is shown that the large $B$ multiskyrmions given by rational map ansaetze
can be described within the domain wall approximation, or
as spherical bubble with energy and baryon number density concentrated at its
boundary.
\vglue 0.6cm}
\elevenrm
\baselineskip=14pt
\section{Introduction}
The chiral soliton approach provides a very economical method of description
of baryonic systems with different baryon numbers, starting with few basic 
concepts and ingredients incorporated in the model lagrangian \cite{1,2}.
The latter is truncated lagrangian of effective field theories
widely used for description and explanation of low-energy meson and baryon 
interactions \cite{3}. Baryons or baryonic systems appear within this approach 
as quantized solitonic solutions of equations of motion, characterized by the
so called winding number or topological charge.
If the concept of topological soliton models is accepted and baryons 
are skyrmions indeed, then
it is clear why there is isospin in the Nature: the number of generators of
the $SU(2)$ isospin group, $3$, coincides with the number of the space 
dimensions, thus allowing for the correlation between $SU(2)$ chiral fields
and space coordinates resulting in appearence of topological solitons.

As it was found numerically, the chiral field configurations of lowest
          energy possess different topological 
properties - the shape of the mass and $B$-number distribution - for different 
values of $B$. It is a sphere for $B=1$ hedgehog \cite{1}, a torus for $B=2$ 
\cite{4}, tetrahedron for $B=3$, cube for $B=4$ \cite{5}, and higher
          polyhedrons for greater baryon numbers \cite{5,6,7}.
A paradoxical feature of the whole approach is that the baryons/nucleons
individuality is absent in the lowest energy static configurations (note,
that any of known lowest energy configurations can be made of a number of
slightly deformed tori). It is believed that the 
standard picture of nuclei should appear when non-zero modes motion - vibration, 
breathing - are taken into account. To get idea about the relative position of
states with different quantum numbers - spin, isospin, flavor, 
$SU(3)$ representation, etc., it is necessary to calculate zero-mode quantum 
corrections to the energy of baryonic system. 
Corrections of this kind have been calculated first for the configurations of
"hedgehog" type \cite{8}, later for configurations with axial symmetry 
\cite{9,10} and for more general configurations, in $SU(2)$ \cite{11} 
as well as $SU(3)$ cases \cite{12,13}.

The chiral soliton approach provides the concept of nuclear matter different
from that widely accepted as being constructed from separate nucleons, only.
To find the "smoking gun" for this unusual concept it is necessary to find
some states which cannot be made of separate nucleons because, e.g. of the
Pauli exclusion principle. The simplest possibility is to consider the
$B=2$ system where the Pauli principle strictly and unambiguosly forbids
definite sets of quantum numbers for the system consisting of separate 
nucleons.

Here we discuss first the $SU(2)$ case (next Section) where supernarrow
low lying dibaryons were predicted \cite{14} and estimate their 
          electromagnetic
decay width. Further the $SU(3)$ extention of the chiral soliton model is
considered and estimates for spectra of multibaryons with flavour
(strangeness, charm or bottom quantum number) made previously are extended
up to highest baryon numbers where the necessary theoretical information
on multiskyrmions is available \cite{7}. A simplified model for large $B$ 
multiskyrmions given by rational maps \cite{15} is presented which allows
us to establish the 
connection with the domain wall or bag approximation (Section 4).
The technical details necessary for calculations are available in literature,
and here some of them given in Appendices,
where several statements valid for any chiral solitons are proved and useful
expressions for $SU(2)$ skyrmion tensors of inertia still lacking in the 
literature are presented.\\
\section{Narrow dibaryons below $NN\pi$ threshold}
     The topological chiral solitons (skyrmions) are classical
          configurations of chiral fields incorporated in unitary matrix
          $U \in SU(2)$ or $SU(3)$ and characterized by topological, or
          winding number identified with baryon number $B$. The classical
          energy (mass) of these configurations $M_{cl}$ is found usually by 
          minimization of energy functional depending on chiral fields. As
          any extended object skyrmions possess also other characteristics
          like moments of inertia $\Theta$ (tensors of inertia in general
          case, see Appendix A), mean square radii of mass and baryon
          number distribution, etc.
The quantization of zero modes of chiral solitons allows to obtain the 
spectrum of states with different values of quantum numbers: spin, isospin,
strangeness, etc. \cite{8}-\cite{13}.
This approach allows for quite reasonable description of varyous 
properties of baryons, nucleons and hyperons, therefore, it is of interest to 
consider predictions of the models of this kind for baryonic systems with
$B \geq 2$. The energy of $SU(2)$ quantized states with axial symmetry can 
be presented as \cite{9,10}
$$ E=M_{cl}+ \frac{I(I+1)}{2\Theta_I} + \frac{J(J+1)}{2\Theta_J} +
\frac{J_3^2}{2B^2 \Theta_3} \biggl(1-\frac{\Theta_3}{\Theta_I}-B^2
\frac{\Theta_3}{\Theta_J}  \biggr)  \eqno (1) $$
Here $I$ and $J$ are isospin and spin of the system, $J_3$ is body fixed
$3-d$ component of the angular momentum which can be considered as an 
          additional internal quantum number of the system. $B=n$ -
          azimuthal winding number for the lowest energy axially
          symmetrical configurations.
          This formula, being obtained rigorously from the model
          lagrangian \cite{9,10}, is very transparent in its physical meaning.
          The technical details beginning with known lagrangian of the
          Skyrme model, expressions for $M_{cl}$, tensors of inertia and 
          some other formulas can be found in Appendix A.

The (generalized) axial symmetry of the configuration with $B=2$ leads to 
definite constraint on the body-fixed
3-d components of the isospin and angular momentum: $J_3 = -n I_3=-nL$ \cite{9,10}. 
As a consequence of this, the states with $I=1$ and $J=0$, or $I=0, \; J=1$, 
and also
$I=J=1$ should have $I_3^{bf}=J_3^{bf}=L=0$. Therefore, the last term
in $(1)$ proportional to $J_3^{bf\,2}$ is absent in these cases. Since the
          parity of configuration equals to $P=(-1)^L$ \cite{10}, all
          states mentioned above have positive parity.
The state with $I=0, \; J=2$ can
have $I_3^{bf}=J_3^{bf}=0$ as well as, e.g., $I_3^{bf}=L=1, \; J_3^{bf}=-2$.
          At large 
$B$ by special reasons (see Appendix A) also only the first two terms in 
          $(1)$, $\sim I(I+1)$ and $\sim J(J+1)$ are important in quantum 
          correction to the energy.

As it was noted long ago \cite{9} the quantum correction for the deuteron-
          like state 
with $I=0,\, J=1$, $E^{rot}_d=1/\Theta_J(B=2)$ is by about $\sim 30 Mev$
smaller than that of the "quasi-deuteron" state with $I=1,\,J=0$,
$E^{rot}_{d'} = 1/\Theta_I(B=2)$. This takes place for all known variants of the
model, without any tuning of parameters, therefore, it can be considered
as intrinsic property of chiral soliton models originating from effective
field theories.
Further investigations of nonzero modes of two-nucleon system, not all but
many of them, have shown that the binding energy of the deuteron can be
reduced to $\sim 6 Mev$ \cite{16} if it is considered as a difference
between states with deuteron and quasideuteron quantum numbers.
Previously and here we consider the differences of energies of quantized
states because they are free of many uncertainties due, e.g., to unknown
loop corrections to the masses of skyrmions (see \cite{17,18} and discussions
below).

According to expression $(1)$ dibaryons are predicted decoupled from $2$-
nucleon channel as a consequence of the Pauli principle \cite{14}. For example, the 
state with isospin $I=J=1$, positive parity and the energy below the threshold 
for the decay into $NN\pi$ with $E^{rot}_D = 1/\Theta_J(B=2) + 1/\Theta_I(B=2)$.
This dibaryon cannot be seen in nucleon-nucleon interactions directly,
but can be observed in reaction $NN \rightarrow NN\gamma\gamma$, where one 
photon is necessary to produce $D$ and the second one appears from the 
decay of $D$: e.g., $pp \rightarrow D^{++}\gamma \rightarrow pp \gamma\gamma.$ 
The chiral soliton models predict the state $D$ with isospin $I=J=1$ at
the energy about $50-60 \, Mev$ above the $NN$ threshold \cite{14}. 

In the paper \cite{10} it was shown that the states with even sum $I+J$ 
($0,\, 2,$ etc.) and positive parity are forbidden by the Finkelstein-
Rubinstein type constraints which appear as
a consequence of requirement that the configuration can be presented as a
system of two unit hedgehogs at large relative distances, and these unit
skyrmions possess fermionic properties. It means, that the configurations
which cannot be considered as consisting of two nucleons, were ignored in
\cite{10}.
Opposite to this, we abandoned this requirement \cite{14}.
It should be noted also that the state with $I=0,\, J=2$ which was forbidden in 
\cite{10} can be in fact the $^3D_2$ state of two nucleons and should not be
forbidden by $FR$ - constraint. Therefore, this particular case should be
analyzed more carefully.

It is possible to estimate the width of the radiative decay $D\rightarrow
          NN \gamma $. Electromagnetic nucleon formfactors can be
          described quite well within Skyrme soliton model in wide
          interval of momentum transfers \cite{19}, reasonable agreement
          with data takes place for deuteron and $2N$ - system \cite{10},
          therefore, one can expect reasonable predictions for systems
          with greater baryon numbers or with unusual properties. The 
          dimensional estimate of narrow dibaryon decay width was made
          in \cite{14} providing the
lower bound for the decay width of few $eV$.
To make more realistic estimate one can consider magnetic type transition
$D \to NN\gamma$ or $d\gamma$.
The amplitude of the direct process due to magnetic dipole 
transition can be written as
          $$ M_{D \to NN\gamma} = ie\; \tilde{\mu}_{D\to NN}\;
          \epsilon_{ikl} F_{ik} \Psi^D_l 
\phi_1^\dagger \phi_2 \eqno (2) $$
where $\tilde{\mu}$ is the value of the transition magnetic moment, we assume that 
$\mu$ is of the order of $\mu_p$, $F_{ik}=e_iq_k-e_kq_i$ - the electromagnetic 
field strength, $\Psi^D_l, \, \phi_1$ and $\phi_2$ are the wave functions of the 
dibaryon and nucleons. For the width of such direct decay we obtain then
          $$ \Gamma_{D \to NN\gamma} = \alpha \Delta M^2
          \frac{\tilde{\mu}^2_{D\to NN}} {945 \pi^2}
 (\Delta/M)^{7/2} \eqno (3) $$
which is numerically less than $0.1 \,eV$ for $\mu \sim \mu_p-\mu_n \simeq
4.7/(2M_N)$, $\Delta = M_{D} -2M$ is the energy
release, or the maximal energy of emitted photon. This estimate agrees with 
          that made previously \cite{14}, but the final state interaction
          could increase it by several orders of magnitude. 

To take it into account roughly one should consider the transition $D \to d'$
where $d'$ is spin zero quasideuteron, or $D^+\to d$.
At this point the important statement is that the isovector magnetic 
          transition operator for any skyrmion is connected simply
          with its mixed, or interference tensor of inertia 
          $\Theta^{int}_{ab}$. This statement, known for
          some particular cases \cite{8,10} is proved in Appendix $B$ for 
          arbitrary skyrmions and for any type of chiral soliton models:
$$\tilde{\mu}_i^a = -{1\over 2} R^{aj} (A) \Theta^{int}_{jk} O^k_i (A'), 
          \eqno (4) $$
          where $R^{aj}=D^1_{aj}=Tr(A^\dagger\tau^aA\tau^j)/2$ and 
          $O^k_i$ are the final rotation matrices, $a$ is isotopical 
(octet in $SU(3)$) index, and for electromagnetic
interaction we should take $a=3$. $\Theta^{int}_{jk}$ is presented in Appendix A.

For configurations with generalized axial symmetry, as well as for several
          known multiskyrmions, only diagonal elements of
          $\Theta^{int}$ are different from zero, moreover, only the
          $(33)$ component remains in the case of axial symmetry, and we have
$$\tilde{\mu}_i^3 = -{1\over 2} R^{33}(A) \Theta^{int}_{33} O^3_i (A'), 
          \eqno (5) $$
$\Theta^{int}_{33} = 2\Theta^I_{33} = 14.8 \,Gev^{-1} $ for $B=2$ and accepted 
values of model parameters, see also {\bf Table 1} below.
To get numerical values of the transition magnetic moments one should calculate
the matrix elements of rotation matrices between the wave functions of
          initial and final states which are equal in terms of final
          rotation matrices $D^I_{I_3,L}$, see e.g. \cite{20}
$$\Psi^D_{I,I_3;J,J_3}=\sqrt{\frac{2I+1}{8\pi^2}} D^I_{I_3L}
 \sqrt{\frac{2J+1}{8\pi^2}} D^J_{J_3, -2L}, \eqno(6)  $$
and we have for $D$ state $I=J=1$, $L=0$, for the final $d'$
state also $I=1$, and $J=0$.
Since $R^{33}=D^1_{00}$ the isotopical part of the matrix element for
          $D\to d'$ transition is proportional to
 $$  <D^1_{I_30}D^1_{00}D^1_{I_30}>= \int D^1_{I_30}D^1_{00}D^1_{I_30}d\nu=
          C^{1,I_3}_{1,0;1,I_3}C^{1,0}_{1,0;1,0}/3. \eqno (7)$$ 
The Clebsch-Gordan coefficient $C^{1,0}_{1,0;1,0}=0$, therefore $D\to d'$
          transition magnetic moment equals to zero for all states including
          $D^{++}$ and $D^0$, not only for $D^+\to d'^+$ which is 
          trivial, and this is a consequence of symmetry
          property of rotator wave function with $L=0$.

For the transition $D^+ \to d\gamma$ the isotopical part of the matrix
          element differs from zero,
 $ <D^1_{0,0}D^1_{00}D^0_{00}>=1/3 $, but the angular momentum part
          proportional to $<D^1_{J_30}D^1_{00}D^1_{J_30}>$ equals to zero
          again.
However, the decay $D^+ \to np$ is possible as a result of isospin violation in the second 
order in electromagnetic interaction, due to virtual emission and reabsorbtion
of the photon and due to isospin violation by the mass difference of $u$ and 
$d$ quarks.
The order of magnitude estimate of the width of this decay due to the virtual 
electromagnetic process is
$$\Gamma_{D\to pn} \simeq \alpha^2 {M \over 4\pi} \sqrt{\Delta/M} \eqno(8)$$
which is about $\sim 1 KeV$. It should be noted here that for the components of
$D$ with charge $+2$ or $0$ the decay into $pp$ or $nn$ final states is
forbidden strictly, due to the rigorous conservation of angular momentum and
the Pauli principle.

For transition $D^{++} \to pp\gamma$, $D^0 \to nn\gamma$ as well as
          $D^+\to (pn)_{I=1}\gamma $ the 
isoscalar magnetic momentum operator gives nonzero contribution.
The corresponding matrix element
$$M_{D \to d'\gamma} = i e\; \tilde{\mu}^0_{D\to d'}\,\epsilon_{ikl} F_{ik} 
\Psi^D_l\Psi^{d'\dagger}  \eqno (9) $$ 
The approximate relation takes place for rational map parameterization:
$$ \tilde{\mu}^0_3 \simeq J_3 \frac{B<r_0^2>}{ 3\Theta^J}, \eqno (10) $$
    $<r_0^2>$ is mean square radius of $B$-number distribution.
$(10)$ coincides with result of \cite{8} for $B=1$ and is close to result
of \cite{10} for $B=2$. The derivation of $(10)$ valid for rational maps 
          parametrization of skyrmions, will be given elsewhere. 
          The coefficient before
$J_3$ in $(10)$ depends remarkably weak on baryon number, as can be established
from {\bf Table 1}. However, numerically $(10)$ gives about twice smaller
result for $B=1$ for parameters we take here, than in \cite{8}. 
As a result we have:
$$ \tilde{\mu}^0_{D \to d'} \simeq \frac{2<r_0^2>}{3\Theta^J},\eqno (11) $$
For the decay width we obtain then
$$ \Gamma_{D \to d'\gamma} =\alpha \frac{4\tilde{\mu}^2_{D\to d'} \Delta^3}{3} 
\eqno (12)$$
Numerically, $\tilde{\mu}_{D\to d'} \simeq 0.35 \,Gev^{-1}$, and from $(12)$ 
 $\Gamma_{D\to d'\gamma} \sim 0.3 \,Kev\,(\Delta/60\,Mev)^3$.
           The same estimate is
valid for the decay rate of $D^+ \to np\gamma$ with $np$-system in isospin
$I=1$ state.

The experimental evidence for the existence of narrow dibaryon $D$ in reaction
 $pp \to pp \gamma \gamma$ has been obtained in Dubna \cite{21}, although
these data have not been confirmed in the Uppsala bremsstrahlung experiment
\cite{22}.
Even more clear indications for the existence of low lying dibaryons have been 
obtained in experiment at Moscow meson factory in the reaction 
$pd \rightarrow pX$ \cite{23}.
The checking and confirmation of these results in its importance is 
comparable with a discovery of new 
elementary particle. The absence of such states would provide definite
restrictions on applicability of the chiral soliton approach and effective 
field theories.

It should be noted that within the model there is a problem of the lowest state
with $I=J=0$ which should be lower than the deuteron-like state. Therefore,
deuteron should decay into this $(0,0)$ state and a photon, but two-nucleon
system in singlet $^1S_0$ state could not decay because $0\to 0$ transition is
forbidden for electromagnetic interaction. 
          The loop corrections to the energy of states, or Casimir energy
          \cite{16}, are different
for states which can go over into two nucleons, and for states which cannot.
Their contribution can change the relative position of these states and shift the $(0,0)$
state above the deuteron, but very nontrivial calculation should be made to
check this.

Some low-lying states with strangeness are also predicted, which cannot decay 
strongly due to the parity and isospin conservation in strong interactions
\cite{14}. For example, the dibaryon with strangeness $S=-2$, $I=0$,
$J=1$ and positive parity has energy by $\sim 0.17 \,Gev$ above 
$\Lambda\Lambda$ threshold \cite{24}, and it cannot decay into two 
$\Lambda$-hyperons because of the Pauli principle, and into $\Lambda\Lambda\pi$
final state by isospin conservation. Therefore, the width of electromagnetic
decay of such state should be not more than few tenths of $Kev$.
It is, of course, a special case. Other possible states with flavour
$s,\, c$ or $b$ will be discussed in the next section.

The masses of neutron rich light nuclides, such as tetra-neutron, 
          sexta-neutron, etc. can be estimated using formula $(1)$. 
For multineutron state with $I=B/2$ the rotation energy
$E^{rot}=B(B+2)/(8\Theta_I)$, and such nuclides are predicted well 
above the threshold for the strong decay into final nucleons. With
increasing baryon numbers the energies of neutron rich states with fixed
          difference $N-Z$ become lower,
so, their width can be quite small. The mass difference of states with
          isospin $I$ and ground states with $I=0$ (for even $B$) equals
          to $\Delta E(B,I) = I(I+1)/(2\Theta_{I,B})$. For such pairs of
          nuclei as $^8Li -^8Be$, $^{12}B-^{12}C$ and $^{16}N-^{16}O$
          it equals to $\Delta E(B,1)=1/\Theta_{I,B}$ and decreases with
          increasing $B$, i.e. atomic number, both theoretically (see 
          {\bf Table 1.} below) and
          according to data. For $B=16$ this difference equals to
          $10.9\,Mev$ in comparison with theoretical value of
          	  $15.8\,Mev$ which is not bad for such a crude model.\\

\section{Flavoured multibaryons}
Another characteristic prediction is that of multibaryons with different 
values of flavours, such as strangeness, charm or bottom quantum numbers. 
The bound state approach of multiskyrmions with different flavours is the
adequate method to calculate the binding energies of states with
quantum numbers $s,\;c$ or $b$. The so called rigid oscillator model
is the most transparent and controlable version of this method \cite{25}.
The references to pioneer papers can be found  also in \cite{26}.
The binding energies of flavoured states are
predicted smaller than binding energies of ordinary nuclei - for strangeness
quantum numbers, and greater - for charm or bottom quantum numbers.
Here I present the main results on flavoured multibaryons following the
papers \cite{26} and extended to higher values of baryon numbers.

To quantize the solitons in $SU(3)$ configuration space,  in the
spirit of the bound state approach to the description of strangeness,
we considered the collective coordinates  motion of the 
meson fields incorporated into the matrix $U \in SU(3)$, see Appendix A:
$$ U(r,t) = R(t) U_0(O(t)\vec{r}) R^{\dagger}(t), \qquad
 R(t) = A(t) S(t), \eqno (13) $$
where $U_0$ is the $SU(2)$ soliton embedded into $SU(3)$ in the usual 
way (into the left upper corner), $A(t) \in SU(2)$ describes $SU(2)$ rotations,
$S(t) \in SU(3)$ describes 
rotations in the ``strange", ``charm" or ``bottom" directions, 
and $O(t)$ describes rigid rotations in real space.
$$ S(t) = exp (i{\cal D}(t)),\qquad
 {\cal D} (t)=\sum_{a=4,...7} D_a(t) \lambda_a, \eqno (14)$$
$\lambda_a$ are Gell-Mann matrices of the $(u,d,s)$, $(u,d,c)$ or $(u,d,b)$
$SU(3)$ groups. The $(u,d,c)$ and $(u,d,b)$ $SU(3)$ 
groups are quite analogous to
the $(u,d,s)$ one. For the $(u,d,c)$ group a simple redefiniton of 
hypercharge should be made. For the $(u,d,s)$ group,
 $D_4=(K^++ K^-)/\sqrt{2}$, $D_5=i(K^+-K^-)/\sqrt{2}$, etc.,
for the $(u,d,c)$ group $D_4=(D^0+\bar{D}^0)/\sqrt{2}$, etc.

The angular velocities of the isospin rotations are defined in the standard 
way: $ A^{\dagger} \dot{A} =-i\vec{\omega}\vec{\tau}/2.$
We shall not consider here the usual space rotations explicitly because the
          corresponding moments of inertia for baryonic systems $(BS)$ are
          much greater than isospin
moments of inertia, see {\bf Table 1.}, and for lowest possible values of angular momentum $J$ the
corresponding quantum correction is either exactly 
zero (for even $B$), or small.

The field $D$ is small in magnitude, at least,
 of order $1/\sqrt{N_c}$, where $N_c$ is the number of colours in $QCD$.
Therefore, an expansion of the matrix $S$ in $D$ can be made safely. 
To the lowest order in field $D$ the Lagrangian of the model $(A1)$
can be written as
$$L=-M_{cl,B}+4\Theta_{F,B} \dot{D}^{\dagger}\dot{D}-\biggl[\Gamma_B 
\bar{m}_D^2+ \tilde{\Gamma}_B (F_D^2-F_\pi^2) \biggr] D^{\dagger}D -
 i{N_cB \over 2}(D^{\dagger}\dot{D}-\dot{D}^{\dagger}D), \eqno(15)$$
$\bar{m}_D^2 = (F_D^2/F_\pi^2) m_D^2-m_\pi^2$.
Here and below $D$ is the doublet $K^+, \, K^0$ ($D^0, \, 
D^-$, or $B^+,\,B^0$).
$\Theta_F$ is the moment of inertia for the rotation into the ``flavour" 
direction ($F=s,\,c,$ or $b$, the index $c$ denotes the 
charm quantum number, except in $N_c$):
$$\Theta_{F,B} = {1\over 8} \int (1-c_f) \biggl[ F_D^2+{1\over e^2}
\biggl((\vec{\partial}f)^2 +s_f^2 (\vec{\partial}n_i)^2\biggr)\biggr] d^3r,
\eqno (16) $$
where $f$ is the profile function of skyrmion, $F_D$ is flavour decay constant, i.e. decay constant of kaon, or $D$-
meson, or $B$-meson,
$$\Gamma_B = {F_\pi^2 \over 2}\int (1-c_f) d^3r \eqno (17) $$
The mass term contribution to static soliton energy is connected with $\Gamma$
due to relation $M.t.=m_\pi^2 \Gamma/2$.
The quantity $\tilde{\Gamma}_B$ enters when flavour symmetry breaking in flavour decay
constants is taken into account:
$$\tilde{\Gamma}_B = {1\over 4}\int c_f \bigl[(\vec{\partial}f)^2 +
s_f^2 (\vec{\partial}n_i)^2 \bigr] d^3r. \eqno (18) $$
It is connected with other calculated quantities via relation:
$$\tilde{\Gamma}= 2(M_{cl}^{(2)}/F_\pi^2-e^2\Theta_F^{Sk}), $$
where $M_{cl}^{(2)}$ is second order contribution into static mass of the
soliton, $\Theta_F^{Sk}$ is Skyrme term contribution into flavour moment of
inertia.
          The contribution proportional to $\tilde{\Gamma}_B$ in $(15)$ is
          suppressed in comparison with the term $\sim \Gamma$
by the small factor $\sim F_D^2/m_D^2$, and is more important for strangeness.
The term proportional to $N_cB$ in $(15)$ arises from the Wess-Zumino term
in the action and is responsible for the difference of
the excitation energies of strangeness and antistrangeness 
(flavour and antiflavour in general case) \cite{25,26}.

Following the canonical quantization procedure the Hamiltonian of the 
system, including the terms of the order 
of $N_c^0$, takes the form \cite{25}:
$$H_B=M_{cl,B} + {1 \over 4\Theta_{F,B}} \Pi^{\dagger}\Pi + \biggl(\Gamma_B 
\bar{m}^2_D+\tilde{\Gamma}_B(F_D^2-F_\pi^2)+\frac{N_c^2B^2}{16\Theta_{F,B}} 
\biggr) D^{\dagger}D +i {N_cB \over 8\Theta_{F,B}}
(D^{\dagger} \Pi- \Pi^{\dagger} D). \eqno (19) $$
$\Pi$ is the momentum canonically conjugate to variable $D$
which describes the oscillator-type motion of  
the $(u,d)$ $SU(2)$ soliton in $SU(3)$ configuration space. After the 
diagonalization which can be done explicitly \cite{25}, the normal-ordered 
Hamiltonian can be written as
$$H_B= M_{cl,B}+\omega_{F,B}a^{\dagger}a+\bar{\omega}_{F,B}b^{\dagger}b
 + O(1/N_c), \eqno (20) $$
with $a^\dagger$, $b^\dagger$ being the operators of creation of strangeness,
i.e., antikaons, and antistrangeness
(flavour and antiflavour) quantum number, $\omega_{F,B}$ and 
$\bar{\omega}_{F,B}$ being the 
frequences of flavour (antiflavour) excitations. $D$ and $\Pi$ are connected
with $a$ and $b$ in the following way \cite{25}:
$$ D^i= (b^i+a^{\dagger i})/\sqrt{N_cB\kappa_{F,B}}, \qquad
\Pi^i = \sqrt{N_cB\kappa_{F,B}}(b^i - a^{\dagger i})/(2i) \eqno (21) $$
with
$ \kappa_{F,B} =[ 1 + 16 (\bar{m}_D^2 \Gamma_B+(F_D^2-F_\pi^2)\tilde{\Gamma}_B
 \Theta_{F,B})/ (N_cB)^2 ]^{1/2}. $
For the lowest states the values of $D$ are small:
$ D \sim \bigl[16\Gamma_B\Theta_{F,B}\bar{m}_D^2 + N_c^2B^2 \bigr]^{-1/4}, $
and increase, with increasing flavour number $|F|$ like $(2|F|+1)^{1/2}$.
As was noted in \cite{25}, deviations of the field $D$ from the vacuum 
decrease with increasing mass $m_D$, as well as with increasing number of 
colours $N_c$, and the method works for any $m_D$ (and also for charm and 
bottom quantum numbers).
$$ \omega_{F,B} = N_cB(\kappa_{F,B} -1)/(8\Theta_{F,B}), \qquad
 \bar{\omega}_{F,B} = N_cB(\kappa_{F,B} +1)/(8\Theta_{F,B}).\eqno (22)$$
As was observed in \cite{26}, the difference
$\bar{\omega}_{F,B}-\omega_{F,B} = N_cB/(4\Theta_{F,B})$ coincides, to the 
leading order in $N_c$ with the expression obtained in the collective 
coordinates approach \cite{24}.

The flavor symmetry breaking $(FSB)$ in the flavour decay constants, 
          i.e. the fact that $F_K/F_\pi 
\simeq 1.22$ and $F_D/F_\pi=1.7 \pm 0.2$ (we take $F_D/F_\pi=1.5$ and
$F_B/F_\pi=2$) leads to the increase of the flavour excitation 
frequences, in better agreement with data for charm and bottom. 
It also leads to some increase of the binding energies of $BS$ 
\cite{26}. 

\vspace{2mm}
\begin{center}
\begin{tabular}{|l|l|l|l|l|l|l|l|l||l|l|l|}
\hline
 $B$  &$M_{cl}$& $\Theta_F^{(0)}$ & $\Theta_I$&$\Theta_{I,3}$&$\bar{\Theta}_J$&
$\Gamma$&$\tilde{\Gamma}$&$<r_0>$&$\omega_s$&$\omega_c$ &$\omega_b$\\
\hline
$1$&$1.702$&$2.05$&$5.55$&$5.55$&$5.55$&$4.80$&$15 $&$2.51$&$0.309$ &$1.542$ 
&$4.82$\\
\hline
$2$&$3.26$ &$4.18$&$11.5$&$7.38$&$23 $ &$9.35$&$22$&$3.46$ &$0.293$ &$1.511 $
&$4.76$\\
\hline
$3$&$4.80 $&$6.34$&$14.4$&$14.4$&$49$&$14.0$&$27$&$4.10$&$0.289$ &$1.504 $
&$4.75$\\
\hline
$4$&$6.20 $&$8.27$&$16.8$&$20.3 $&$78$&$18.0$&$31 $&$4.53$&$0.283$ &$1.493$
&$4.74 $\\
\hline
$5$&$7.78$ &$10.8$&$23.5 $&$19.5$&$126$&$23.8$&$35 $&$5.10$&$0.287 $&$1.505$
&$4.75 $\\
\hline
$6$&$9.24 $&$13.1$&$25.4$&$27.7$&$178$&$29.0$&$38 $&$5.48$&$0.287 $&$1.504 $
&$4.75 $\\
\hline
$7$&$10.6 $&$14.7$&$28.9$&$28.9$&$220$&$32.3$&$43$&$5.72$&$0.282 $&$1.497 $
&$4.75 $\\
\hline
$8$&$12.2 $&$17.4$&$33.4$&$31.4$&$298$&$38.9$&$46$&$6.15$&$0.288 $&$1.510 $
&$4.79 $\\
\hline
\hline
$9$&$13.9 $&$20.5$&$37.7$&$37.7$&$375$&$46$&$47$&$6.49$&$0.291 $&$1.517$
&$4.77 $\\
\hline
$12$&$18.4 $&$28.0$&$48.5$&$48.5$&$636$&$64 $&$54$&$7.31$&$0.294$&$1.526$
&$4.79 $\\
\hline
$13$&$19.9 $&$30.5$&$52.0$&$52.0$&$737$&$70 $&$57$&$7.5?$&$0.288$&$1.497$
&$4.70 $\\
\hline
$14$&$21.5 $&$33.6$&$56.1$&$56.1$&$865$&$78 $&$59$&$7.85$&$0.299$&$1.536$
&$4.80 $\\
\hline
$16$&$24.5 $&$38.9$&$63.1$&$63.1$&$1107$&$91$&$63 $&$8.31$&$0.301 $&$1.543$
&$4.81$\\
\hline
$17$&$25.9 $&$41.2$&$66.1$&$66.1$&$1219$&$96 $&$65 $&$8.48$&$0.300 $&$1.542$
&$4.81$\\
\hline
$22$&$33.7 $&$56.0$&$84.2$&$84.2$&$2027$&$135 $&$73 $&$9.55$&$0.308 $&$1.560$
&$4.84$\\
\hline
$32^*$&$49.1$&$86.7$&$118$&$118$&$4154$&$218 $&$87 $&$11.3$&$0.319$ &$1.585$
&$4.84$\\
\hline
\end{tabular}
\end{center}
\vspace{2mm}
{\baselineskip=10pt
\tenrm
{\bf Table 1.} Characteristics of the bound states of skyrmions
with baryon numbers up to $B=22$. The classical mass of solitons $M_{cl}$ is
in $GeV$, moments of inertia $\Theta_F,\,\Theta_I$, $\Theta_{I,3}$,
and $\Theta_J$ as well as $<r_0>$, $\Gamma$ and $\tilde{\Gamma}$ - in 
$GeV^{-1}$, the excitation frequencies for flavour $F$, $\omega_F$ in $GeV$. 
$<r_0>=\sqrt{r^2_B}$ with $\Theta_J$ defines the value 
of isoscalar magnetic momentum of multiskyrmion. For larger baryon numbers, 
          beginning with $B=9$, calculations are made using rational maps
          $(RM)$ ansatz. For $B=32$ it 
was assumed that the ratio ${\cal I}/B^2 = 1.28$ as for $RM \; B=22$ skyrmion.
The external parameters of the model are
$F_{\pi}=186 \, MeV, \; e=4.12$. The accuracy of calculations is better
than $1\%$ for the masses and few $\%$ for other quantities. }\\
\vspace{2mm}

$\bar{\Theta}_J$ shown in {\bf Table 1} is $1/3$ of the trace of corresponding
tensor of inertia, see Appendix A.
As it can be seen from {\bf Table 1} the flavour excitation energies increase
again for $B =22$, and the important property of binding
becomes weaker for largest $B$. It can be, however, the artefact of the
$RM$ approximation discussed in the next Section. In particular, for 
          rational maps solitons with $B \geq 9$ we take as moment of
          inertia $\Theta_I$  and $\Theta_{I,3}$ $1/3$ of the trace of 
          corresponding tensor of inertia, see Appendix A.

For large value $F_D/F_\pi =\rho_D$ and mass $m_D$, the following
approximate formula for the flavor excitation frequences can be obtained:
 $$\omega_{F,B}\simeq \tilde{m}_D\biggl(1-2\frac{\Theta^{Sk}_{F,B}}
  {\rho_D^2\Gamma_B}\biggr)-\frac{N_cB}{2\rho_D^2\Gamma_B} \eqno(23) $$
          with $\tilde{m}_D^2=m_D^2+F_\pi^2\tilde{\Gamma}_B/\Gamma_B$.
It is clear from $(23)$ that, first, $\omega$'s are smaller than masses
          of mesons $m_D$, i.e. the binding takes place always, and it is
          to large degree due to the contribution of the Skyrme term into
          the flavour inertia $\Theta_F^{Sk}$. When $\rho_D \to \infty$,
          $\omega_F \to m_D$. Since the ratios $\tilde{\Gamma}_B/\Gamma_B$
          decreases with increasing $B$ and $\Theta_{F,B}/\Gamma_B$
increases when $B$ increases from 1 to 4 - 7, the energies $\omega_{F,B}$
decrease for these $B$-numbers, therefore, it leads to increase of binding
 of flavoured mesons by $SU(2)$ solitons with increasing $B$ up
 to $B \sim 4 - 7$. However, for $B=22$ and $32$ the ratio
 $\Theta_{F,B}/\Gamma_B$ is smaller than for $B=1$, and, indeed,
 the $\omega$'s are the same and even larger than for $B=1$.

\vspace{2mm}
\begin{center}
\begin{tabular}{|l|l|l|l||l|l|l||l|l||l|}
\hline
 $B$ & $\Delta \epsilon_{s=-1}$&
$\Delta\epsilon_{c=1}$ &$\Delta \epsilon_{b=-1}$&$\Delta \epsilon_{s=-2}$&
$\Delta\epsilon_{c=2}$ &$\Delta \epsilon_{b=-2}$ \\
\hline
$2$&$-0.047$&$-0.03$&$0.02$&$-0.053$&$-0.07 $&$0.02 $ \\
\hline
$3$&$-0.042$&$-0.01 $&$0.04$&$-0.036$&$-0.03 $&$0.06 $ \\
\hline
$4$&$-0.020 $&$0.019$&$0.06$&$-0.051$&$ 0.022$&$0.10 $ \\
\hline
$5$&$-0.027 $&$0.006$&$0.05$&$-0.063$&$0.001$&$0.08 $ \\
\hline
$6$&$-0.019 $&$0.016$&$0.05$&$-0.045$&$0.023$&$0.10 $ \\
\hline
$7$&$-0.016 $&$0.021$&$0.06$&$-0.041$&$0.033$&$0.11 $ \\
\hline
$8$&$-0.017 $&$0.014$&$0.02$&$-0.040$&$0.021$&$0.03 $ \\
\hline
\hline
$9$& $-0.023 $&$0.005$& $0.03$& $-0.10$&$-0.003$&$0.06$ \\
\hline
$12$&$-0.021 $&$0.003$& $0.02$& $-0.09$&$-0.004$&$0.04$ \\
\hline
$17$&$-0.027 $&$-0.013$&$0.00$& $-0.11 $&$-0.03$&$-0.00$ \\
\hline
$22$&$-0.034 $&$-0.028$&$-0.03$&$-0.14$&$-0.06$&$-0.03 $ \\
\hline
\end{tabular}
\end{center}
\vspace{2mm}
{\baselineskip=10pt
\tenrm
{\bf Table 2.}
The binding energy differences $\Delta \epsilon_{s,c,b}$ are the changes
of binding energies of lowest $BS$ with flavour $s,\,c$ or $b$ and isospin
$I=T_r+|F|/2$ in comparison
with the usual $u,d$ nuclei, for the flavour numbers $S=-1,\, -2 $,
 $c=1,\, 2$, $b=-1$ and $-2$ (see Eq. (24)). The $SU(3)$
multiplets are $(p,q)=(0, 3B/2)$ for even $B$ and $(p,q)=(1,(3B-1)/2) $
for odd $B$.}\\

The binding energies differences between flavoured multibaryons and ordinary
nuclei in the rigid oscillator approximation are given by the formula:
$$\Delta\epsilon_{B,F}=|F|\biggl[\omega_{F,1}-\omega_{F,B} -
\frac{3(\kappa_{F,1}-1)}{8\kappa_{F,1}^2\Theta_{F,1}} -
\frac{T_r(\kappa_{F,B}-1)}{4\kappa_{F,B}\Theta_{F,B}} -
\frac{(|F|+2)(\kappa_{F,B}-1)^2}{8\kappa_{F,B}^2\Theta_{F,B}}\biggr],\eqno(24)$$
and the lowest $SU(3)$ multiplets are considered with isospin of flavourless
component $T_r=0$ for even $B$ and $T_r=1/2$ for odd $B$.
This formula is correct for $|F|=1$ and for any $|F|$ if the baryon number is
large enough to ensure the isospin balance.

The values of $\Delta \epsilon$ shown in {\bf Table 2.} should be considered as 
an estimate.
They illustrate the restricted possibilities of $RM$ approximation for
large $B$ multiskyrmions.

Isosinglet $BS$, in particular those with $|F|=B$ are of special interest.
As it was argued in \cite{26} such states do not belong to the lowest possible
$SU(3)$ irreps, they should have $T_r=|F|/2$.
It makes sense to calculate the difference of binding energy of such state and
the minimal state $(p^{min},q^{min}$ with zero flavour which we identify
with usual nucleus (ground state):
$$\Delta\epsilon_{B,F}=|F|\biggl[\omega_{F,1}-\omega_{F,B} -
\frac{3(\kappa_{F,1}-1)}{8\kappa_{F,1}^2\Theta_{F,1}} +
\frac{(|F|+2)(\kappa_{F,B}-1)}{8\kappa_{F,B}^2\Theta_{F,B}}\biggr]-$$
$$-{1 \over 2\Theta_{T,B}}[|F|(|F|+2)/4-T_r^{mim}(T_r^{min}+1)]
 \eqno (25) $$
where $T_r^{min}=0$, or $1/2$ as before.
\begin{center}
\begin{tabular}{|l|l|l|l||l|l|l||l|l|l||l|}
\hline
 $B$  &$\Delta \epsilon_{s=-1}$& $\Delta \epsilon_{c=1}$& $\Delta\epsilon_{b=-1}$
  &$\Delta \epsilon_{s=-2}$& $\Delta\epsilon_{c=2}$ & $\Delta \epsilon_{b=-2} $
  &$\Delta \epsilon_{s=-3}$& $\Delta\epsilon_{c=3}$ &
$\Delta \epsilon_{b=-3}$ &$\Delta\epsilon_{s=-B}$\\
\hline
$2$& $-$&$-$ & $-$ & $-0.075$&$-0.03 $&$0.02$&$-$ & $-$ &$-$&$-0.07 $\\
\hline
$3$&$0.000$&$0.034$&$0.07 $&$- $&$ -$ &$-$&$-0.08 $&$0.002$&$0.09$&$-0.08 $ \\
\hline
$4$&$- $&$- $&$- $&$-0.047$&$0.030$  &$0.09$& $-$ & $-$ &$-$&$-0.13 $ \\
\hline
$5$&$-0.003 $&$0.032 $&$0.06 $&$-$&$-$&$-$&$-0.06 $&$0.035$&$0.12$&$-0.15$ \\
\hline
$6$&$- $&$- $&$- $&$-0.044$&$ 0.025$  &$0.09$&$-$ & $-$&$-$&$-0.21 $  \\
\hline
$7$&$ 0.000$&$0.040$&$0.07$&$- $&$-$&$-$&$-0.04 $&$0.068$&$0.15$&$-0.20$\\
\hline
$8$&$- $&$- $&$- $&$-0.039$  &$ 0.023$ &$0.03$& $-$ & $-$ & $-$&$-0.28$\\
\hline
$12$&$- $&$- $&$- $&$-0.046$  &$ 0.00 $ &$0.03$& $-$ & $-$ & $-$&$-0.50$\\
\hline
$17$&$ -0.020$&$-0.01$&$-0.00$&$- $&$-$&$-$&$-0.081$&$-0.04$&$-0.01$&$-0.82$\\
\hline
$22$&$- $&$- $&$- $&$-0.073$  &$ -0.06 $ &$-0.06$& $-$ & $-$ & $-$&$-1.3$\\
\hline
$32^*$&$- $&$- $&$- $&$-0.088$  &$-0.11 $ &$-0.13 $& $-$ & $-$ & $-$&$--$\\
\hline
\end{tabular}
\end{center}
{\baselineskip=10pt
\tenrm
{\bf Table 3.} The binding energies differences of lowest flavoured $BS$ with
isospin $I=0$ and the ground state with the same value of $B$ and
$I=0$ or $I=1/2$.
The first $3$ columns are for $|F|=1$, the next $3$ columns -
for $|F|=2$, and the next 3 - for $|F|=3$. The state with the value of
flavour $|F|$ belongs to the $SU(3)$ multiplet with $T_r= |F|/2$.
In the last column the binding energies differences are shown for the
isoscalar electrically neutral states with $S=-B$. For $|F| \geq 3$ all
estimates are very approximate.}\\

According to {\bf Table 3} the total binding energy of state e.g. with $B=22$
and $S=-2$ is by $73 \, Mev$ smaller than that of nucleus $A=22$, so it should 
be well bound.
The model used here is too crude for large values of flavour, and results
obtained can be used only for illustration and as a starting point for
further investigations.
Results similar to those described in this section are obtained also in other
versions of the model \cite{27}, in particular in the quark-meson soliton model 
\cite{28}. For baryon numbers $B=3,4$ estimates of spectra of baryonic systems
with charm quantum number were made in \cite{29} within conventional quark
model. They are in fair agreement with ours.

In the channel with $B=2$ the near-threshold state with strangeness $S=-1$
was observed long ago in the reaction $pp \rightarrow p\Lambda K^+$ \cite{30}
and confirmed recently in COSY experiment \cite{31}. 
Similar near-threshold $\Lambda\Lambda$ state was observed by KEK PS E224 
collaboration \cite{32}.
The Skyrme model explains these near-threshold states
with $B=2$, and predicts similar states for greater values of the $B$-number.
For some values of $B$, beginning with $B \geq 5,6$ such states with several 
units of strangeness can be stable relative to strong interactions.
Due to well known relation between charge, isospin and hypercharge of hadrons,
$Q=I_3+(B+S)/2$, the $BS$ with several units of strangeness can appear as
negatively charged nuclear fragments. For even $B$ and minimal multiplets
$(p,q)= (0, 3B/2)$, strangeness $S=-2I$, and condition when $Q=-1$ fragment
appears first is $-1=S+B/2$, or $-S=B/2+1$. For $B=6$ it is $S=-4$, for $B=8$,
$S=-5$, etc. For odd $B$ the $Q=-1$ state should have strangeness 
$-S=(B-1)/2+1$, i.e. $-3,\;-4$ and $-5$ for $B=5,\;7$ and $9$, etc.

The negatively charged long living nuclear fragment with mass about 
$7.4 \, Gev$ observed in $NA52$ CERN experiment in $Pb\,+\,Pb$ collision 
at the energy of $158\,A\, Gev$ \cite{33}
 can be, within the chiral soliton models,
a fragment with $B=7$ or $6$ and strangeness $S= -4$ or $-5, -6$.
Confirmation of this result as well as searches for other negatively
charged fragments would be of great importance.
For charm or bottom quantum numbers the binding energies are greater, but
to observe such states one needs considerably higher incident energies.\\
\section{Large $B$ multiskyrmions from rational maps in the domain wall 
approximation}
The rational map ansatz for skyrmions proposed in \cite{15} and
widely used now, also in present paper, simplified considerably
the treatment of multiskyrmions, and, at the same time, it leads to the 
picture of multibaryon system at
large $B$ which is, probably, incompatible with a picture for ordinary
nuclei. To clarify this point, we consider here large $B$ multiskyrmions
in some kind of a toy model - in a domain wall approximation, which gives,
however, quite good numerical results for known $RM$ multiskyrmions,
except $B=1,2$.
The energy of skyrmion for rational map ansatz \cite{15} in universal 
units $3\pi^2 F_\pi/e$ is:
$$ M={1 \over 3\pi}\int \biggl\{A_N r^2f'^2 +2Bs_f^2(f'^2+1) + 
{\cal I} \frac{s_f^4}{r^2} \biggr\} dr  \eqno (26) $$
The coefficient $A_N = 2(N-1)/N$ for symmetry group $SU(N)$ \cite{34}.
The quantity ${\cal I}$ for $SU(2)$-case is given in Appendix A, the 
          inequality takes place ${\cal I} \geq B^2 $.
Direct numerical calculations have shown, and the analytical treatment
here supports, that at large $B$ and, hence, large ${\cal I}$ multiskyrmion looks
like a spherical ball with profile equal to $F=\pi$ inside and $F=0$
outside. The energy and $B$-number density of this configuration is concentrated 
at its boundary, similar to the domain walls system considered in \cite{35}
in connection with cosmological problems.

Consider such large $B$ skyrmion within the "inclined step" approximation.
Let $W$ be the width of the step, and $r_0$ - the radius of the 
skyrmion where the profile $f=\pi/2$. $f= \pi/2 - (r-r_o)\pi/W$
for $r_o-W/2 \leq r \leq r_o+W/2 $.
 Note that this approximation describes the usual domain wall energy \cite{35}
with accuracy $\sim 9 \%$.

We write the energy in terms of $W, r_0$, then minimize it
with respect to both of these parameters, and find the minimal value of
energy.
$$ M(W, r_0)={\pi^2 \over W}(B+A_Nr_0^2)+W \biggl(B+
\frac{3 {\cal I}}{8r_0^2}\biggr) \eqno (27) $$
This gives 
$$W_{min}=\pi\biggl[\frac{B+A_Nr_0^2}{B+3{\cal I}/(8r_0^2)}
\biggr]^{1/2} \eqno (28) $$
and, after minimization, $r^2_{0\,min}=\sqrt{3 {\cal I} /(8 A_N)}$.
In dimensional units $r_0= (6{\cal I}/A_N)^{1/4} / (F_\pi e)$.
Since $ {\cal I} \geq B^2 $, the radius of minimized 
configuration grows as $\sqrt{B}$, at least.
$W_{min}=\pi$, i.e. it does not depend on $B$ for any $SU(N)$.
The energy
$$M_{min}\simeq (2B+\sqrt{3A_N {\cal I} /2}) /3 \eqno (29) $$
For $SU(2)$ model $A_N=1$ and the energy $M_{min}= (2B+\sqrt{3{\cal I} /2})/3$
should be compared with the lower bound $M_{LB}=(2B+\sqrt{I})/3$.
The formula gives the numbers for $B=3,..., \, 22$ in remarkably good agreement
with calculation within $RM$ approximation, within $2-3 \%$ \cite{7}.

It is not difficult to calculate the corrections to these expressions,
of relative order $1/B,\; 1/B^2,...$:
$$ M(W, r_0)\simeq {\pi^2 \over W }(B+A_Nr_0^2)+W \biggl[
B(1+\beta)+\frac{3 {\cal I}}{8 r_0^2}(1+\gamma)\biggr], \eqno (30) $$
$\beta=\pi^2/(12B), \; \gamma=(2\pi^2+17)/\sqrt{24\cal{I}}$.
$$M_{min} \simeq [2B(1+\beta/2)+\sqrt{3{\cal I}/8}(1+\gamma/2)] \eqno (31) $$
However, the first order correction in $W$ does not improve the
description of masses, and summation of all terms seems to be required.

So, we see that a very simple approximation provides a confirmation of
a picture from numerical calculation of $RM$ skyrmions as a two-phase
object, a spherical ball with profile $f=\pi$ inside and $f=0$ outside,
and a fixed width envelope  with fixed surface energy density,
          $\rho_M = M/(4\pi r_0^2)\simeq (2B+\sqrt{3{\cal I} /2 })/ (12\pi
r_0^2)$. At large $B$ $\rho_M \to Const$, but average mass density 
over the volume $ \to 0$.

Consider also the influence of the mass term which gives the contribution
$$ M.t. =\tilde{m} \int r^2(1-cos\,F) dr , \eqno (32) $$
$\tilde{m} = 8m_\pi^2/(3\pi F_\pi^2 e^2) $. For strangeness, charm, or bottom
the masses $m_K$, $m_D$ or $m_B$ should be inserted instead of $m_\pi$.
In the "inclined step" approximation we obtain:
$$ M.t. \simeq \tilde{m} \biggl[{2\over 3}r_o^3 + O(W^2 ) \biggr] \eqno (33)$$
In view of this structure of the mass term it makes no influence on the
width of the step $W$ in lowest order, but the dimension of the soliton
$r_o$ becomes smaller:
$$ r_o \to r_o - \tilde{m} \frac{r_o^2(B+A_Nr_o^2)}{4\pi B}. \eqno (34) $$
As it was expected from general grounds, dimensions of the soliton decrease
with increasing $\tilde{m}$.
However, even for large value of $\tilde{m}$ the structure of multiskyrmion at 
large $B$
remains the same: the chiral symmetry broken phase inside of the spherical
wall where the main contribution to the mass and topological charge is 
concentrated. The behaviour of the energy density for $B=22$ at different 
values of $\mu$ is shown in Fig. The value of the mass density inside of
the ball is defined completely by the mass term with $1-c_f = 2$.
The baryon number density distribution is quite similar, with only difference
that inside the bag it equals to zero.
It follows from these results that $RM$ approximated multiskyrmions cannot
model real nuclei at large $B$, probably $B > 12 -20$, and configurations like
skyrmion crystals may be more valid for this purpose.

Besides the simple one-shell configurations considered in \cite{7,15} and here,
multishell configurations can be of interest. Some examples of two-shell
configurations with $B=12,\,13,\,14$ have been considered recently \cite{36}.
The profile $f=2\pi$ at $r=0$ for such configurations and decreases to $f=0$
for $r\to \infty$. We can also model such two-shell configuration in the
domain-wall, or spherical bag approximation with a result
$$M\simeq (2B_1+\sqrt{3{\cal I}_1/2})/3+ (2B_2+\sqrt{3{\cal I}_2/2})/3,
\eqno (35) $$
with total baryon number $B=B_1+B_2$. The profile $f$ decreases from $2\pi$
to $\pi$ in the first shell, and from $\pi$ to $0$ in the second. The radii of 
both shells should
satisfy the condition $r_0^{(2)} \geq r_0^{(1)}+ W$, so external shell 
should be large enough, with baryon number $B_2$ of several tens, at least.
Since the ratio ${\cal I}/B^2$ is greater for smaller $B$, the energy $(35)$
is greater than the energy of one-shell configuration considered before.
Calculations performed in \cite{36} also did not give results better than
for one-shell configurations.
However, more refined consideration would be of interest.
Observation concerning the structure of large $B$ multiskyrmions made here
can be useful in view of possible cosmological applications of Skyrme-type
models.

\section{Concluding remarks}
Here we have restricted ourselves with the Skyrme model and its straightforward 
extentions. However, many of the result are valid in other variants of the 
model: in the model with solitons stabilized by the explicit vector ($\omega$)
meson, or stabilized by the baryon number density squared, in the chiral
perturbation theory, etc, see discussion in \cite{14}b. The $B=2$ 
torus-like configuration has
          been obtained within these models, as well as in the chiral
          quark-meson model \cite{28},
and it would be of interest to check if there are also multiskyrmions with
$B \geq 3$.

We did not discuss a special classes of $SU(3)$ skyrmions, $SO(3)$ solitons
and the problems of their observation, as well as $SU(3)$ skyrmion 
molecules. The discussion of these topics can be found in several papers 
of \cite{12,13}. Some new solutions which are not $SU(2)$ embeddings in
$SU(3)$ or $SU(n)$ have been found in \cite{34}.
          
To conclude, the study of some processes, also at intermediate energies which, 
to some extent, are out of fashion now, can provide a very important check of 
fundamental principles and concepts of the elementary particles theory 
including the confinement of quarks and gluons. 
The confirmation of the predictions of the chiral soliton approach would 
provide qualitatively new understanding of the origin of nuclear forces. 
If the existence of low energy radiatively decaying dibaryons is reliably
established, it will change the long standing beliefe that nuclear matter 
fragments should consist necessarily of separate nucleons bound by their 
interactions. Therefore, the confirmation and checking of the results of
experiments on dibaryons production, as well as of production of fragments
of flavoured matter is extremely important. It would be possible at 
accelerators of 
moderate energies, like COSY (Juelich, FRG), KEK (Japan), Moscow meson factory 
(Troitsk, Russia), ITEP (Moscow), and some others.
The production of multistrange states, as well as states with charm or bottom
quantum numbers, is possible in heavy ion collisions, and also on accelerators
like Japan Hadron Facility to be built in the near future.

The multiple flavour production realized in the production of flavored
multibaryons possible, e.g., in heavy ion collisions,
demands higher energy, of course, but multiple interaction processes and
normal Fermi motion of nucleons inside of nuclei make effective thresholds
much lower \cite{37}. Studies of such flavoured multibaryons production would 
allow more complete and reliable checking of the model predictions. 

We note finally that the low energy dibaryons have been obtained recently
in \cite{38} using the quantization procedure different from our. 
          
The work is supported by RFBR grant 01-02-16615,
UK PPARC grant PPA/V/ S/1999/00004 and presented in part at the International
seminar Quarks-2000, Pushkin, Russia, May 2000.
\\

{\large \bf Appendix A. Inertia tensors of multiskyrmions.}\\

The lagrangian density of the $SU(2)$ Skyrme model is given by 
$$ {\cal L} = -\frac{F_\pi^2}{16} Tr L_\mu L_\mu + {1 \over 32e^2} 
Tr G_{\mu \nu }^2 + \frac{F_\pi^2 m_\pi^2}{16}Tr (U+U^\dagger -2),  
\eqno (A1) $$
$L_\mu = \partial_\mu U U^\dagger $ is left chiral derivative, $L_\mu =
i L_{\mu,k}\tau_k$, $\tau_k$ are Pauli matrices.
$G_{\mu\nu}=\partial_\mu L_\nu - \partial_\nu L_\mu$ - strengh of the
chiral field. The Wess-Zumino term present in the action has been
discussed in details in \cite{13}, and we shall omit this discussion here.

First we give the expression for the energy of $SU(2)$ - skyrmion as
a function of profile $F$ and unit vector 
 $\vec{n}$, which is especially useful in some cases. Using definition
$U=c_f+is_f\vec{n}\vec{\tau}$, and relation
$$L_{\mu ,k} L_{\nu ,k}=\partial_\mu f\partial_\nu f+
s_f^2\partial _\mu \vec{n}\partial _\nu \vec{n}, \eqno (A2)$$
we obtain
$$M_{stat}= \int \Biggl\{ {F_\pi^2 \over 8}[(\vec{\partial}f)^2+s_f^2
(\vec{\partial}
n_i)^2] + {s_f^2 \over 4e^2} \biggl[2[\vec {\partial }f\vec {
\partial }n_i]^2+s_f^2 [\vec {\partial }n_i \vec {\partial }n_k]^2 \biggr] +
 \rho_{M.t.} \Biggr\} d^3r. \eqno (A3) $$

For the Ansatz based on rational maps the profile
$F$ depends only on variable $r$, and components of vector $\vec{n}$ - on
angular variables $\theta, \, \phi$. 
$n_x=(2\, Re\, R)/(1+|R|^2) ,\; n_y=(2\, Im\, R)/(1+|R|^2) ,\; n_z =(1-|R|^2)/
(1+|R|^2)$, where $ R$ is a rational function of variable $z=tg(\theta/2)
exp(i\phi)$ defining the map from $S^2 \to S^2$.
In this case the gradients of functions
 $F$ and
$\vec{n}$ are orthogonal (recall that $\vec {\partial}_r = \vec{n}_r \partial_r
+\vec{n}_\theta \partial_\theta/r +\vec{n}_\phi \partial_\phi/(r s_\theta)$ ¨
$\vec{n}_r=\vec{r}/r=(s_\theta c_\phi, \, s_\theta s_\phi, \, c_\theta)$, 
$\vec{n}_\theta= (-c_\theta c_\phi, \,
-c_\theta s_\phi,\, s_\theta)$,  $\vec{n}_\phi=(s_\phi, \, -c_\phi, \, 0))$
and $[\vec{\partial}f\vec{\partial}n_1]^2=f'^2 (\vec{\partial}n_1)^2$, etc.
Taking into account relations
$$n_3^2[\vec{\partial}n_2\vec{\partial}n_3]^2=n_1^2[\vec{\partial}n_1
\vec{\partial}n_2]^2, 
\qquad
n_3^2[\vec{\partial}n_1\vec{\partial}n_3]^2=n_2^2[\vec{\partial}n_1
\vec{\partial}n_2]^2,  \eqno (A4)$$
one can present $(A3)$ as
$$M_{stat}= \int \Biggl\{ {F_\pi^2 \over 8}[ (f')^2+s_f^2(\vec {\partial}
n_i)^2] + {s_f^2 \over 2e^2} \biggl[ f'^2(\vec {\partial }n_i)^2+
s_f^2 [\vec {\partial }n_1 \vec {\partial }n_2]^2/n_3^2 \biggr] + 
\rho_{M.t.} \Biggr\} d^3r. \eqno (A5) $$
Usually the notation is introduced
$${\cal I} = {1 \over 4\pi}\int r^4\frac{[\vec{\partial}n_1\vec{\partial}n_2]^2}
{n_3^2} d\Omega = {1 \over 4\pi} \int \Biggl(\frac{(1+|z|^2)}{(1+|R|^2)}
{|dR| \over |dz|}\Biggr)^4 \frac{2i dz d\bar{z}}{(1+|z|^2)^2}, \eqno (A6) $$
and using the equation
$$ \int r^2(\vec{\partial}n_k)^2 d\Omega =2 \int r^2\frac{|[\vec{\partial}n_1
\vec{\partial}n_2]|}{|n_3|} d\Omega =2 \int \frac {2i dR d\bar{R}}{(1+|R|^2)^2}=
8\pi{\cal N} \eqno (A7) $$
obtain finally
$$M_{stat}= 4\pi \int \Biggl\{ {F_\pi^2 \over 8} (f'^2r^2+2 s_f^2 {\cal N}) + 
{s_f^2 \over 2e^2} \biggl[2 f'^2{\cal N}+
s_f^2 {\cal I}/r^2 \biggr] + r^2 \rho_{M.t.} \Biggr\} dr. \eqno (A8) $$
To find the minimal energy configuration at fixed
${\cal N}=B$ one minimizes ${\cal I}$, and then finds the profile $F(r)$ by 
minimizing energy $(A8)$.

To quantize zero modes one uses ansatz
$U(t,r)=A(t) U(O_{ik}(t)r_k)A^\dagger(t)$, and evident relation
$$\partial_tU = \dot{U}= \dot{A}U(\vec{r}(t))A^\dagger + AU(\vec{r}(t))
\dot{A}^\dagger + \dot{r}_i(t)A \partial_iU(\vec{r}(t))A^\dagger, \eqno (A9) $$
where $r_i(t)=O_{ik}(t)r_k$ is body-fixed coordinate.
 
Angular velocities of spatial (or orbital) rotations are introduced
according to:
$$\dot{r}_i = \dot{O}_{ik}r'_k = \dot{O}_{ik}O^{-1}_{kl}r_l(t)=
-\epsilon_{ilm}\Omega_m r_l(t) $$
and integration is performed in coordinate system bound to soliton (body-fixed).

The rotation, or zero-mode energy of $SU(2)$ skyrmions as a function of 
angular velocities is
$$E_{rot}= {1\over 2} \Theta^I_{ab}\omega_a\omega_b+
\Theta^{int}_{ab}\omega_a\Omega_b+
{1\over 2} \Theta^J_{ab}\Omega_a\Omega_b .\eqno (A10) $$
The isotopical tensor of inertia for arbitrary $SU(2)$ skyrmion is:
$$\Theta^I_{ab}=\int s_f^2\Biggl\{(\delta_{ab}-n_an_b)\Biggl({F_\pi^2 \over
4}+\frac{(\vec{\partial} f)^2}{e^2}\Biggr) +\frac{s_f^2}{e^2}\partial_ln_a
\partial_l n_b \Biggr\} d^3r. \eqno (A11) $$
For the RM ansatz trace of this tensor of inertia is
$$ \Theta^I_{aa}(RM)=4\pi \int s_f^2\Biggl\{{F_\pi^2\over 2} +{2\over e^2}
\Biggl(f'^2 +{\cal N}\frac{s_f^2}{r^2} \Biggr)\Biggr\} r^2dr. \eqno (A12) $$

The orbital inertia tensor gives contribution to the energy
$ \Theta^J_{ab} \Omega_a\Omega_b/2$, and for arbitrary configuration
using the same notations is given by:
$$\Theta^J_{ab}=\int \Biggl\{{F_\pi^2 \over 4}(
\partial_if\partial_kf+s_f^2
\partial_i\vec{n}\partial_k\vec{n}) +{s_f^2 \over e^2}\biggl[
\partial_if\partial_kf(\vec{\partial}n_l)^2+(\vec {\partial }f)^2
\partial_i\vec{n}\partial_k\vec{n}-$$
$$-\partial_if\partial_lf\partial_l\vec{n}\partial_k\vec{n}-
\partial_kf\partial_lf\partial_l\vec{n}\partial_i\vec{n}+
s_f^2[(\vec{\partial}n_l)^2\partial_i\vec{n}\partial_k\vec{n}-
(\partial_i\vec{n}\partial_l\vec{n})(\partial_k\vec{n}\partial_l\vec{n})]
\biggr]\Biggr\} \epsilon_{i\alpha a}\epsilon_{k\beta b}r_{\alpha}r_\beta d^3r.
\eqno (A13) $$
This expression can be simplified for $RM$ ansatz:
$$\Theta^J_{ab}=\int s_f^2 \Biggl\{\Biggl[{F_\pi^2 \over 4}+{f'^2 \over e^2}+
{s_f^2 \over e^2}(\vec{\partial}n_l)^2\Biggr]
\biggl[(\vec{\partial}n_l)^2(r^2\delta_{ab}-r_ar_b)-\partial_a\vec{n}
\partial_b\vec{n}r^2\biggr] - $$
$$ -{s_f^2 \over e^2}
\biggl[(\partial_i\vec{n}\partial_k\vec{n})(\partial_i\vec{n}\partial_k\vec{n})
(r^2\delta_{ab}-r_ar_b)-r^2
(\partial_a\vec{n}\partial_l\vec{n})(\partial_b\vec{n}\partial_l\vec{n})\biggr]
\Biggr\}d^3r  \eqno (A14) $$
It allows to obtain easily the trace of the inertia tensor $\Theta^J_{aa}$.
$$ \Theta^J_{aa}(RM)=4\pi \int s_f^2\Biggl\{{F_\pi^2\over 2}
{\cal N} +
{2\over e^2}\Biggl(f'^2 {\cal N}+{\cal I}\frac{s_f^2}{r^2}\Biggr) \Biggr\}
r^2dr. \eqno (A15) $$

It is easy to establish the inequality for traces of isotopical and orbital
tensors of inertia:
$$ \Theta^J_{aa} - B \Theta^I_{aa}={8\pi \over e^2}({\cal I}-B^2)
\int s_f^4 dr \geq 0, \eqno (A16) $$
since ${\cal I} \geq B^2,\, {\cal N}=B$.
The interference (mixed) tensor of inertia which defines also the isovector
part of the magnetic transition operator equals to:
$$\Theta^{int}_{ab}= \int s_f^2\Biggl\{\Biggl[{F_\pi^2 \over 4}+{1 \over e^2}
\bigl[(\partial_\nu f)^2+s_f^2(\partial_\nu \vec{n})^2\bigr]\Biggr]\partial_i n_l-$$
$$-{1\over e^2}(\partial_i f\partial_\nu f+s_f^2 
\partial_i\vec{n}\partial_\nu\vec{n})\partial_\nu n_l
\Biggr\} n_k\epsilon_{kla} \epsilon_{i\alpha b} r_\alpha d^3r. \eqno (A17) $$
The components of spatial angular velocities interfere with components
$\omega_1, \, \omega_2, \, \omega_3$ of angular velocities of rotation in
configuration space, only.

Numerically the components of the mixed tensor of inertia are much smaller
than those of isotopical, or orbital tensor of inertia, except special cases
of "hedgehogs" when
$-\Theta^{int}=\Theta^I=\Theta^J$, and axially symmetrical configurations 
when for the $3-d$ components of inertia relations hold
$-\Theta^{int}_{33}= n \Theta^I_{33}= \Theta^J_{33}/n $. 

Note, that most general formulas for tensors of inertia are presented 
here for the first time. For the case of $RM$ configurations they differ in
some details from those given in the literature.\\

{\large \bf Appendix B. Electromagnetic transition operators.}\\

 Here we prove, for completeness, in general form some statements concerning 
isovector (octet in $SU(3)$ case) vector charge and isovector magnetic 
momentum operator.

There is the following connection between isovector current and isospin
generator
$$ V_{0,a}={1 \over 2} Tr (A^\dagger \lambda_aA\lambda_b) I_b^{bf} =
R_{ab}(A)I_b^{bf}, \eqno (B1)$$
where the isospin generator in body-fixed (connected with soliton)
coordinate system is
$$I_b^{bf}=\partial L^{rot}(\omega,\Omega) / \partial\omega_b. \eqno (B2)  $$
$a,b=1,2,3$ for $SU(2)$-model, and $a,b=1,...8$ for $SU(3)$-model.
To prove this consider ansatz
$$U=e^{-i\alpha_a\lambda_a/2} A(t)U_0A^\dagger (t)e^{i\alpha_a\lambda_a /2}
 \eqno (B3) $$
The vector Noether current is a coefficient before derivative of the probe
function, $\partial_\mu \alpha$. In the lowest order in $\alpha$
we obtain for the chiral derivative:
$$U^\dagger \partial_0U =A\bigl[U_0^\dagger A^\dagger(\dot{A}-i\dot{\alpha}
A/2)U_0 - A^\dagger(\dot{A}-i\dot{\alpha}A)\bigr]A^\dagger \eqno (B4) $$
Using the definition of angular velocities of rotation in configuration
space $\omega_a$, we obtain
$$ A^\dagger \dot{A}-iA^\dagger \dot{\alpha}A/2=-{i \over 2}\lambda_b
(\omega_b+R_{ab}(A)\dot{\alpha}_a), \eqno (B5) $$
where real orthogonal matrix
$$R_{ab}(A)={1\over 2}Tr(A^\dagger \lambda_a A\lambda_b). \eqno (B6) $$
Since the dependence on $\dot{\alpha}$ 
reduces to simple addition to angular velocity according to
$(B5)$, formula $(B1)$ follows immediately.

According to the well known relation,
$$Q= B+I_3/2 =B+V_{0,3}/2 \eqno (B7) $$
the baryonic (topological) charge and $3-d$ component of the isospin
generator contribute to the charge of the quantized skyrmion.

We prove also that there is simple connection between isovector (octet for
$SU(3)$ model) magnetic momentum operator of the skyrmion and mixed
(interference) tensor of inertia.
Note first that the lagrangian of arbitrary chiral model, not only Skyrme model,
because of Lorentz invariance can be presented as a sum, with some coefficients,
of contributions of the type:
$${\cal L}_{M,N}= Tr\bigl(U^\dagger\dot{U}MU^\dagger\dot{U}N- U^\dagger\partial_kUM
U^\dagger\partial_kUN\bigr), \eqno (B8) $$ 
where $M$ and $N$ are some matrices. E.g., for the second order term
$M=N=1$. 
The contribution into rotational energy, proportional to $\Omega$, $\omega$, 
which comes from the first term in $(B8)$ and defines mixed or interference 
tensor of inertia is (see $(A9)$ above):
$$\Theta^{int}_{ab}\omega_a\Omega_b=\int Tr\bigl(U_0^\dagger A^\dagger \dot{A}U_0 -
A^\dagger\dot{A}\bigr)\tilde{M}U_0^\dagger\partial_kU_0\tilde{N}\dot{r}_k d^3r
+ \biggl(M\to N\biggr) ,\eqno (B9) $$
$\tilde{M} = A^\dagger MA, \tilde{N}= A^\dagger NA$.
Or,
$$\Theta^{int}_{ab}=-{i\over 2}\epsilon_{bjk}\int r_j(t)Tr(U_0^\dagger
\lambda_aU_0-\lambda_a)\tilde{M}U_0^\dagger \partial_k U_0\tilde{N} d^3r
+\biggl(M\to N \biggr) \eqno(B10) $$
$r_j(t)$ and $\partial_k$ are body-fixed here.
From the second term in expression $(B8)$ we obtain for the spatial components 
of the vector current:
$$V_k^a= {i\over 2}Tr\bigl(U_0^\dagger A^\dagger \lambda_aAU_0-A^\dagger
\lambda_a A\bigr)\tilde{M}U_0^\dagger\partial_kU_0\tilde{N} +\biggl(M\to N
\biggr) \eqno (B11)$$
Taking into account that
$A^\dagger\lambda_a A= R_{ab}(A)\lambda_b$, $R_{ab}={1\over 2}TrA^\dagger
\lambda_a A\lambda_b $ and $\partial_k = O_{lk} \partial_l^{bf}$,
$$V_k^a= {i\over 2}R_{ab} O_{lk} Tr\bigl(U_0^\dagger \lambda_b U_0-
\lambda_b\bigr)\tilde{M}U_0^\dagger\partial_l U_0\tilde{N} +\biggl(M\to N
\biggr) \eqno (B12)$$
By definition
$$\mu^a_i = {1\over 2} \epsilon_{ijk}\int r_j V_k^a d^3r, \eqno (B13)$$
or
$$\mu^a_i = {i\over 4} \epsilon_{ijk}R_{ab}(A)O_{qk} O_{pj}\int r_p(t) 
Tr(U_0^\dagger\lambda_bU_0-\lambda_b)\tilde{M}U_0^\dagger 
\partial_q U_0\tilde{N} +\biggl(M \to N\biggr). \eqno (B14) $$
Taking into account that
$$\epsilon_{ijk} O_{pj}O_{qk} = \epsilon_{pql} O_{li}  $$
we obtain the desired relation between components of the magnetic momentum
operator and mixed tensor of inertia in the body-fixed coordinate system:
$$\mu^a_i = -{1\over 2} R_{ab}(A) \Theta^{int}_{bl}O_{li}. \eqno (B15)$$
In some particular cases this relation was used previously \cite{8,10}.

For the transition matrix elements calculations it is necessary to average
this expression over wave functions of some initial and final states, see
Section 2.\\

{\elevenbf\noindent References}
\vglue 0.2cm

\vspace{1cm}
{\elevenbf Figure caption.}\\

The mass density distribution of the rational map multiskyrmion
with $B=22$ as a function of the distance from
center of skyrmion for different values of mass in 
the chiral symmetry breaking term . 

a) pion mass in the mass term, b) kaon mass, c) $D$-meson mass,
the mass density is devided by $10$.
                   

\begin{thebibliography}{50}
\tenrm\baselineskip=12pt
\bibitem{1} T.H.R. Skyrme, Proc.Roy.Soc. A260, 127 (1961);
   Nucl.Phys. 31, 556 (1962)
\bibitem{2} E. Witten, Nucl. Phys. B223, 433 (1983)
\bibitem{3} J. Gasser, H. Leutwyler, Nucl. Phys. B250, 465 (1985);
          U-G.Meissner, COSY News, No 7, 6 (1999)
\bibitem{4} V.B. Kopeliovich, B.E. Stern, JETP Lett. 45, 203 (1987);
 NORDITA Preprint 89/34 (1989); J.J.M. Verbaarschot, Phys.Lett. 195B, 235 (1987)
\bibitem{5} E. Braaten, S. Townsend, L. Carson, Phys.Lett. B235, 147 (1990)
\bibitem{6} R.A. Battye, P.M. Sutcliffe,
 Phys.Rev.Lett. 79, 363 (1997); hep-th/0012215 (2000)
\bibitem{7} P.M. Sutcliffe, Talk at the International Seminar Quarks-2000, 
Pushkin, May 2000.
\bibitem{8} G.S. Adkins, C.R. Nappi, E. Witten, Nucl.Phys. B228, 552 (1983);
\bibitem{9} V.B. Kopeliovich, Sov.J. Nucl.Phys. 47, 549 (1988) [Yad.Fiz. 
 47, 1495 (1988)]
\bibitem{10} E. Braaten, L. Carson, Phys.Rev. D38, 3525 (1988)
\bibitem{11} L. Carson, Nucl.Phys. A535, 479 (1991); T.S. Walhout, Nucl.Phys. A547, 423 (1992); P. Irwin, Phys.Rev. D61, 114024 (2000)          
\bibitem{12} G. Guadagnini, Nucl.Phys. B236, 35 (1984);
 A.P. Balachandran, F. Lizzi, V.G.J. Rodgers, A. Stern, Nucl. Phys. B256, 
          525 (1985); B. Schwesinger, H. Weigel, Phys. Lett. B267, 438 (1991);
V.B. Kopeliovich, Phys.Lett. B259, 234 (1991); Sov.J.Nucl.Phys. 51, 151 (1990);          
C.L. Schat, N.N. Scoccola, Phys.Rev. D62, 074010 (2000)          
\bibitem{13} V.B. Kopeliovich, JETP 85, 1060 (1997);
Nucl. Phys. A639, 75c (1998); V.B. Kopeliovich, B.E. Stern, 
W.J. Zakrzewski, Phys. Lett. B492, 39 (2000)
          \bibitem{14} V.B. Kopeliovich, Phys.Atom.Nucl. 58, 1237 (1995);
          ibid. 56, 1084 (1993) [Yad.Fiz. 56, 160 (1993)]
\bibitem{15} C. Houghton, N. Manton, P. Suttcliffe, Nucl. Phys. B510, 507
 (1998)
\bibitem{16} R.A. Leese, N.S. Manton, B.J. Schroers, Nucl. Phys. B442, 228
          (1995)
\bibitem{17} B. Moussalam, Ann. of Phys. (N.Y.) 225, 264 (1993)
\bibitem{18} H. Walliser, Phys. Lett. B432, 15 (1998);
N. Scoccola, H. Walliser, hep-ph/9805340, Phys. Rev. D58, 094037 (1998)
\bibitem{19} G. Holzwarth, Z.Physik, A356, 339 (1996); 
          Nucl.Phys. A666, 24 (2000)
\bibitem{20} L.D. Landau, E.M. Lifshitz, Theoretical physics, Vol. 3
Quantum mechanics, par. 58, 110, Nauka, Moscow, 1974
\bibitem{21} The DIB-2$\gamma$ collaboration, JINR preprint E1-96-104 (1996);
S.B. Gerasimov, nucl-th/9712064; nucl-th/9812077
\bibitem{22} H. Calen et al, Phys.Lett. B427, 248 (1998)
\bibitem{23} E.S. Konobeevsky, M.V. Mordovskoi, S.L. Potashev et al,
Izv. Ross. Acad. Nauk, 62, 2171 (1998);
L.V. Filkov, V.L. Kashevarov, E.S. Konobeevski et al, 
nucl-ex/9902002; hep-ex/9904003, Phys.Rev. C61, 044004 (2000) 
\bibitem{24} V.B. Kopeliovich, B. Schwesinger, B.E. Stern, Nucl.Phys.
A549, 485 (1992)          
\bibitem{25} I.R. Klebanov, K.M. Westerberg, Phys.Rev. D53, 2804 (1996); 
ibid. D50, 5834 (1994); V.B. Kopeliovich, JETP Lett. 67, 896 (1998), 
hep-ph/9805296          
\bibitem{26} V.B. Kopeliovich, W.J. Zakrzewski, JETP Lett. 69, 721 (1999);
Eur. Phys. J. C 18, 369 (2000)          
\bibitem{27} J.P. Garrahan, M. Schvellinger, N.N. Scoccola, Phys.Rev. 
          D61, 014001 (2000); C.L. Schat, N.N. Scoccola, Phys.Rev. 
          D61, 034008 (2000)
\bibitem{28} J. Segar, M. Sripriya, M. Sriram, Phys. Lett. B342, 201 (1995); 
V.B. Kopeliovich, M.S. Sriram, Yad.Fiz. 63, 552 (2000)
\bibitem{29} B.F. Gibson, C.B. Dover, G. Bhamathi, D.R. Lehman,
Phys.Rev. C27, 2085 (1983)          
\bibitem{30} J.T. Reed et al, Phys.Rev. 168, 1495 (1968)                     
\bibitem{31} D. Grzonka, K. Kilian, Nucl.Phys. A639, 569c (1998); 
COSY-TOF collaboration, COSY News, No 4, 1 (1999)          
\bibitem{32} J.K. Ahn et al, AIP Conf. Proc. 412, 923 (1997); Nucl.Phys. A639,
379c (1998)          
\bibitem{33} S. Kabana et al, J. Phys. G23, 2135 (1997)
\bibitem{34} T. Ioannidou, B. Piette, W.J. Zakrzewski, J.Math.Phys. 40, 6353
(1999); ibid. 40, 6223 (1999)          
\bibitem{35} Ya.B. Zeldovich, I.Yu. Kobzarev, L.B. Okun, ZhETF 67, 3 (1974)          
\bibitem{36} N.S. Manton, B.M.A.G. Piette, hep-th/0008110          
\bibitem{37} V.B. Kopeliovich, Phys.Rept. 139, 51 (1986)          
\bibitem{38} T. Krupovnickas, E. Norvaisas, D.O. Riska, nucl-th/0011063  
\end{thebibliography}
\end{document}